
\def \cst{\char'143\char'157\char'156\char'163\char'164}  
\def \ss{\char'140} 
\def \J {{\cal J}}
\def \pd{\partial}
\def \s{\sigma}
\def \b{\beta}
\def \a{\alpha}
\def \g{\gamma}
\def \d{\delta}

\def \l{\lambda}
\def \th{\theta}
\def \ph{\phi}

\def \Ph{\Phi}
\def \p{\pi}
\def \P{\Pi}
\def \D{\Delta}
\def \G{\Gamma}
\def \z {\zeta}
\def \m{\mu}
\font \bigbf=cmbx10 scaled \magstep1
\tolerance=500
\magnification 1200
{\nopagenumbers
\line {\hfil LTH 277}
\line {\hfil January 1992}
\vskip .5in
\centerline{\bigbf    EXACT BOSONIC AND SUPERSYMMETRIC     }
\centerline{\bigbf    STRING BLACK HOLE SOLUTIONS      }
\vskip 1in
\line {\hfil \bf I. Jack, D. R. T. Jones and J. Panvel \hfil}
\vskip 5pt
\centerline {
\it DAMTP, University of Liverpool, Liverpool L69 3BX, UK      }
\vskip 1in
\line {\hfil \bf Abstract \hfil}
\vskip 5pt
We show that Witten's two-dimensional string black hole metric is
exactly conformally invariant in the supersymmetric case. We also
demonstrate that this metric, together with a recently proposed
exact metric for the bosonic case, are respectively consistent with the
supersymmetric and bosonic $\s$-model conformal invariance conditions
up to four-loop order.
\vfill
\eject}
\pageno=1
\line {\bigbf 1. Introduction\hfil}
\vskip 12pt
Lately, intense activity has been devoted to the construction of
conformally invariant theories representing strings propagating in
black-hole type backgrounds, mostly in two dimensions but with some
higher dimensional examples[1,2,3]. This endeavour is not new[2],
         but the recent spate of interest was stimulated by Witten's
observation[1] that the $SL(2,R)/U(1)$ gauged Wess-Zumino-Witten (WZW)
model[4,5] has an interpretation as a
string in a two-dimensional black hole.
The essential point of Witten's insight is to provide us with an
exactly conformally invariant theory with a black hole interpretation,
whereas previous solutions were only valid perturbatively. This opens up
the possibility of investigating the nature of spacetime singularities
in string theory using the techniques of conformal field theory.

In Witten's original analysis, upon gauge-fixing and integrating out the
gauge fields in the $SL(2,R)/U(1)$ gauged WZW model, a non-linear
$\s$-model is obtained with metric and dilaton background fields. The
metric exhibits explicitly the properties associated with a black hole,
such as an event horizon enclosing a spacetime singularity. However,
the metric and dilaton fields only satisfy the $\s$-model conformal
invariance conditions (which are obtained from the $\b$-functions for
the $\s$-model) to lowest order, despite the fact that the
original gauged WZW model was exactly conformally invariant (since it
represents a coset model[6]). This
discrepancy was ascribed to additional contributions arising from the
functional measure when integrating out the gauge field. In a later
analysis[7], the Virasoro operators $L_0$, $\bar L_0$ for the gauged
$SL(2,R)/U(1)$ model were expressed as differential operators. By
identifying these operators with the Laplacian obtained from the string
effective action, proposed exact solutions for the metric and dilaton
background were obtained. This derivation is somewhat indirect and
moreover leaves out of account higher order perturbative corrections
to the string effective action, and hence it is of interest to check
whether this metric/dilaton background does indeed yield a solution to
the conformal invariance conditions beyond one loop. It has recently
been shown[8] that the purported exact solution is in fact consistent
with conformal invariance up to three loops in $\s$-model perturbation
theory. At three loops, issues emerge concerned with scheme dependence
of the conformal invariance conditions, which may be further
illuminated by considering higher orders. Hence one purpose of this
paper is to show that the exact metric and dilaton background given in
Ref. [7] is indeed conformally invariant up to four loops, within
renormalisation scheme ambiguities.

The four-loop calculation is also of interest in the supersymmetric
context. Witten's original string black hole solution was derived for
the bosonic gauged $SL(2,R)/U(1)$ gauged WZW model. However, it is also
possible to construct the supersymmetric extension of this model[9]. By
repeating the analysis of Ref. [7], we shall argue that Witten's
original metric/dilaton background should be exactly conformally
invariant in the supersymmetric case. This is manifestly consistent with
the fact that the $\b$-function for the $N=1$ supersymmetric $\s$-model
is equal to that for the bosonic $\s$-model at one loop and vanishes at
the two[10] and three[11] loop level. However, there is in general a
non-vanishing contribution at four loops[12]. We shall show that although
the four-loop $N=1$ supersymmetric $\b$-function computed using minimal
subtraction fails to vanish for Witten's metric/dilaton background,
nevertheless there exists a renormalisation scheme in which it does.
\vskip 1in
\line {\bigbf 2. The string black hole \hfil}
\vskip 12pt
In this section we briefly recapitulate previous results [1,7] on string
black hole solutions. The Wess-Zumino-Witten (WZW) model[4] may be
regarded as a special case of a nonlinear $\s$-model with an
antisymmetric tensor field background in addition to a metric
background. In general we may gauge a subgroup of the isometry group
of the $\s$-model provided certain conditions are satisfied[13]. For
the WZW model based on the group ${\cal G}$, the isometry group is
${\cal G}\times{\cal G}$, and for a subgroup ${\cal H}\subset{\cal G}$
we can gauge the diagonal subgroup ${\cal H}\times{\cal H}$, providing
a Lagrangian realisation of the coset model ${\cal G}/{\cal H}$. In the
case at hand, we consider the $SL(2,R)$ WZW model and gauge a $U(1)$
subgroup. We can gauge either a vector or an axial realisation
of the $U(1)$ subgroup, corresponding to the symmetry under $g\rightarrow
hghˆ{-1}$ or $g\rightarrow hgh$, respectively. We choose to gauge the
axial subgroup here (gauging the vector subgroup leads to the ``dual''
$\s$-model[7, 14]).
          We parametrise the group by \ss\ss Euler" angles, writing a
generic group element $g$ as
$$g=eˆ{{i \over 2}{\ph_L}{\s_2}}eˆ{{1 \over 2}r{\s_1}}eˆ{{i \over
                2}{\ph_R}{\s_2}} \eqno(2.1)$$

We will consider gauging the group generated by $\s_2$, so that the
local gauge transformations correspond to $\ph_{L,R}\rightarrow
\ph_{L,R} +\a$. This will yield a 2-dimensional black hole of Euclidean
signature. The gauged WZW action takes the form
$$\eqalign{S_{GWZW}=S_{WZW}&+{k \over {2\p}}  \int dˆ2z
               [A(\bar\pd \ph_L +\cosh r\, \bar\pd\ph_R) \cr &
 +(\pd\ph_R+\cosh r\, \pd\ph_L)\bar A - A\bar A(\cosh r+1)]} \eqno(2.2)$$
with the ungauged WZW action $S_{WZW}$ given by
$$S[r,\ph_L,\ph_R] = {k\over{4\p}} \int dˆ2z (\pd r\bar\pd r-\pd\ph_L
\bar\pd\ph_L -\pd\ph_R\bar\pd\ph_R-2\cosh r \,\pd\ph_L\bar\pd\ph_R  )
\eqno(2.3)$$

At this point, following Witten one can pick a gauge by setting  $\ph_L
= -\ph_R =\ph$ and integrate out the gauge fields using their equations
of motion. The result has the form of the 2-dimensional non-linear
$\s$-model action,
$$S={1 \over {4\p\a'}} \int dˆ2x \sqrt\g\, \gˆ{\m\nu} \pd_\m\phˆi \pd_\nu
    \phˆj g_{ij}(\ph)+{1 \over {4\p\a'}}\int dˆ2x \sqrt\g\, D(\ph)Rˆ{(2)}
    \eqno(2.4)$$
where $\g_{\m\nu}$ is a metric on the 2-dimensional worldsheet and
$Rˆ{(2)}$  is the worldsheet Ricci scalar. $\phˆi(\l), i=1,...,N$, can
be regarded as co-ordinates on an N-dimensional manifold with metric $gˆ
{ij}$ and dilaton $D(\ph)$. In the case at hand we have $\{\phˆi\}=\{r,
\ph\}$ with a metric and dilaton given by
$$\eqalign{ dsˆ2 &= {{\a'k}\over4}(drˆ2 + eˆ{2\l(r)}d\phˆ2)\cr
           \l(r) &= \ln\left(2\tanh {r\over 2}\right) \cr
            D(r) &= - \ln \cosh{ r\over 2} + \cst . } \eqno(2.5)$$

In fact the dilaton field does not appear upon naively integrating over
the gauge fields, and its existence is inferred somewhat indirectly.
Moreover, although the metric and dilaton given by Eq.(2.5) satisfy the
$\s$-model conformal invariance conditions at one loop, they fail to do
so at higher order. Since the original gauged $SL(2,R)/U(1)$ WZW model
was exactly conformally invariant, this is a somewhat unsatisfactory
outcome.  This situation can be remedied by following the analysis of
Ref.[7]. The gauge field is parametrised as
$$\eqalign{ A &= \pd \p _L \cr \bar A &= \bar\pd\p _R } \eqno(2.6)$$
with $\p _L=\p _Rˆ*$. Upon making the shift
$$ \ph_L \rightarrow \ph_L + \p _L ,\qquad \ph_R \rightarrow \ph_R
 + \p _R   \eqno(2.7)$$
the action Eq.(2.2) takes the form
$$S_{GWZW} = S_{WZW} + S[\p ] + S[b,c] \eqno(2.8)$$
where
$$\eqalign{S[\p ] &=  {k \over {4\p}} \int dˆ2z \pd\p \bar\pd\p  \cr
  \p  &= \p _L - \p _R} \eqno(2.9)$$
and $S[b,c]$ is a ghost action derived from the Jacobian of the change
of field variables Eq.(2.6), given by
$$S[b,c]=\int dˆ2z ( b\bar\pd c + \bar b \pd \bar c) \eqno(2.10)$$
If we also impose the gauge condition
$$\bar\pd A=\pd\bar A \eqno(2.11)$$
then we can write
$$\p(z,\bar z)=\varphi(z)+\bar\varphi(\bar z). \eqno(2.12) $$
The holomorphic conserved current is given by
$$2k gˆ{-1}\pd g  = i\s_2J_2 +i\s_3{1 \over 2}(J_+ - J_-) + \s_1
  {1 \over 2} (J_+ + J_-) \eqno(2.13)$$
with
$$\eqalign{ J_2 &= (\pd\ph_R + \cosh r \,\pd\ph_L) \cr
  J_\pm &= eˆ{\pm i\ph_R}(\pd r \mp \sinh r\, \pd\ph_L)} \eqno(2.14)$$
The holomorphic part of the energy momentum tensor is given in terms of
the currents through the Sugawara construction [15] as
$$T(z)={1 \over {k-2}}(Jˆ2_2 - J_+J_-) - {k \over 4}(\pd\p )ˆ2
  +b\pd c \eqno(2.15)$$

The physical operators of the theory are defined by requiring that they
commute with the BRST charge
$$Q_{BRST} = \int dz\, c\,(J_2 +{1\over 2} k \pd\p ) + c.c.\eqno(2.16)$$
It is argued in [7] that there is only one propagating degree of
freedom in the 2-dimensional string theory represented by the
$SL(2,R)/U(1)$ models and that this may be taken to be the tachyon
field. The primary fields of the coset conformal field theory represent
the vertex operators for the tachyon field.
Writing the tachyon vertex operator, $V(z,\bar z)$, as
$$V(z,\bar z)=T(r(z,\bar z),\ph_L (z,\bar z),\ph_R(z,\bar z))
         eˆ{iq_R\varphi(z)+iq_L\bar\varphi(\bar z)} \eqno(2.17)$$
we find that the zero modes $Jˆ0_a$ of the $SL(2,R)$ currents in
Eq.(2.12) may be represented as differential operators acting on
$T(r,\ph_L,\ph_R)$ of the form
$$\eqalign{ Jˆ0_2 &=  {\pd \over {\pd\ph_R}} \cr
  Jˆ0_\pm &= eˆ{\pm i\ph_R}\left[{\pd \over{\pd r}} \mp{i\over{\sinh r}}
\,\left({\pd\over{\pd\ph_L}}-\cosh r
  \,{\pd\over{\pd\ph_R}}\right)\right]}\eqno(2.18)$$

Using the condition that $V(z,\bar z)$ in Eq.(2.17) commutes with
$Q_{BRST}$ in Eq.(2.16), we find from Eqs.(2.15), (2.17) and (2.18) that
the Virasoro operators $L_0$, $\bar L_0$ are represented by the operators
$$\eqalign{L_0 &= -{1\over {k-2}}\D_0 - {1\over k} {\pdˆ2\over
  {\pd\phˆ2_R}} \cr \bar L_0 &= -{1\over {k-2}}\D_0
  -{1\over k} {\pdˆ2\over {\pd\phˆ2_L}} }\eqno(2.19)$$
where
$$\D_0 (r,\ph_L,\ph_R)={\pd\over{\pd rˆ2}}+\coth r\,{\pd\over{\pd r}}
  +{1\over{\sinhˆ2 r}}\left({\pdˆ2\over{\pd\ph_Lˆ2}}-2\cosh r\,
  {\pd\over{\pd\ph_L}}{\pd\over{\pd\ph_R}}+{\pdˆ2\over{\pd\ph_Rˆ2}}
  \right)\eqno(2.20)$$
Using the physical state condition
$$(L_0 - \bar L_0 )\,T(r,\ph_L,\ph_R)=0 \eqno(2.21)$$
we can decompose $T(r,\ph_L,\ph_R)$ as
$$\eqalign{T(r,\ph_L,\ph_R) = T(r,\ph)+ \tilde T(r,\tilde\ph) \cr
  \ph = {1\over 2}(\ph_L-\ph_R),\qquad\tilde\ph = {1\over2}(\ph_L+\ph_R)
  }\eqno(2.22)$$
so that acting on $T(r,\ph)$, $L_0$ has the form
$$L_0=-{1\over{k-2}}\left({\pdˆ2\over{\pd rˆ2}}+\coth r\,{\pd\over{\pd r}}
  +{1\over 4}{1\over{\sinhˆ2{r\over2}}}\,{\pdˆ2\over{\pd\phˆ2}}\right)
  -{1\over{4k}}{\pdˆ2\over{\pd\phˆ2}}\eqno(2.23)$$

In the $\s$-model approach to string theory, on the other hand, the
tachyon is described to lowest order by a target-space effective action
of the form
$$S=\int dˆ2\ph eˆ{-2
D}\sqrt g({1\over2}\a'gˆ{ij} \pd_i T \pd_jT -2Tˆ2) \eqno(2.24)$$
The equations of motion derived from this effective action give the
lowest order conformal invariance conditions for the tachyon background
field. The conformal invariance conditions should be equivalent to the
physical state conditions expressed in terms of the Virasoro  operators,
and hence we identify $L_0$ in Eq.(2.23) with the Laplacian derived from
Eq.(2.24), namely
$$L_0 =-{\a'\over{4eˆ{-2
D}\sqrt g}}\pd_ieˆ{-2D}\sqrt g gˆ{ij}\pd_j \eqno(2.25)$$
This identification leads to a solution for $g_{ij}$ and $D$ given by
$$\eqalign{dsˆ2 &={a'\over 4}(k-2)(drˆ2+\bˆ2(r)d\phˆ2) \cr
  D &=-{1\over2} \ln{{\sinh r}\over{\b(r)}} \cr
  \b(r)&=2\left(\cothˆ2{r\over2}-{2\over k}
  \right)ˆ{-{1\over 2}}}\eqno(2.26)$$
This form for the metric and dilaton is claimed to be an exact solution
of the conformal invariance conditions. Before investigating this claim
in detail, we shall repeat the above analysis for the supersymmetric
case.
\vfill
\eject
\line {\bigbf 3.  The Supersymmetric Gauged SL(2,R) WZW model \hfil}
\vskip 12pt
In this section we consider the supersymmetric extension of the gauged
$SL(2,R)$ WZW model. The general supersymmetric WZW model was considered
in Ref. [9] and the gauged $SL(2,R)/U(1)$ version was discussed in
Ref. [16]. The analysis of the previous section may readily be repeated
for this case. The field element in Eq(2.1) is replaced by a superfield
$G(z,\bar z,\th,\bar\th)$ where $\th$, $\bar\th$ are superspace
Grassman coordinates. $G$ has an expansion in terms of components
$$G=g+\th\g+\bar\th\bar\g +\th\bar\th f \eqno(3.1)$$
                                                                and is
parametrised as
$$G(z,\th)=eˆ{{i\over2}\Ph_L\s_2}eˆ{{1\over2}R\s_1}eˆ{{i\over2}\Ph_R
  \s_2} \eqno(3.2)$$
for superfields $\Ph_{L,R},R$.
The supersymmetric gauged WZW action is now of the form
Eqs. (2.2), (2.3), but with
$$\int dˆ2 z \rightarrow \int dˆ2 z dˆ2 \th,\qquad \pd\rightarrow D,
\qquad \bar\pd\rightarrow\bar D,\qquad r\rightarrow R, \qquad
\ph_{L,R}\rightarrow\Ph_{L,R} \eqno(3.3)$$
where the superspace covariant derivatives are given by
$$D={\pd\over{\pd\th}}-\th{\pd\over{\pd z}},\qquad \bar D=
  {\pd\over{\pd\bar\th}}-\bar\th{\pd\over{\pd\bar z}}\eqno(3.4)$$
Following the analysis of Section 2, if we pick the gauge
$$\bar D A=D\bar A \eqno(3.5)$$
we can parametrise the gauge fields as
$$A=D\P_L,\qquad\bar A= \bar D\P_R\eqno(3.6)$$
and then the action reduces to the forn
$$S_{GWZW}=S_{WZW}+S[\P]+S[B,C]\eqno(3.7)$$
with
$$\eqalign{\P&=\P_L-\P_R \cr
S[\P] &= {-k\over{4\p}}\int dˆ2zdˆ2\th D\P\bar D\P \cr
  S[B,C] &= \int dˆ2zdˆ2\th (B\bar DC+ \bar B D\bar C)}\eqno(3.8)$$
where $B$ and $C$ are anti-ghost and ghost superfields corresponding to
the redefinition Eq. (3.6).
If we impose the gauge condition
$$\bar DA=-D\bar A \eqno(3.9)$$
then we can write
$$\P=\Psi+\bar\Psi \eqno(3.10)$$
where
$$\bar D\Psi=D\bar\Psi=0.\eqno(3.11)$$
The holomorphic conserved currents are given by
$$2kGˆ{-1}DG=i\s_2\J_2+\J_1\s_1+i\J_3\s_3 \eqno(3.12)$$
with
$$\J_1={1\over2}(\J_++\J_-),\qquad \J_3={1\over2}(\J_+-\J_-) \eqno(3.13)$$

and
$$\eqalign{\J_2 &= D\Ph_R  +\cosh R\, D\Ph_L \cr \J_\pm &=
  eˆ{\pm i\Ph_R}(DR\mp \sinh R\, D\Ph_L)}\eqno(3.14)$$
The holomorphic part of the superconformal tensor, [9,17], $T(z,\th)$,
   is then given in terms of the  currents by
$$T(z,\th)={1\over k}\left(\J_2D\J_2- {1\over2}
\J_+D\J_- -{1\over2}\J_-D\J_+
  \right)-{2i\over{3kˆ2}}f_{abc}\Jˆa(\Jˆb\Jˆc) -{k\over{4\p}}D\P D\P
  +BDC \eqno(3.15)$$
where $f_{abc}$ are the structure constants for $SL(2,R)$.
We now introduce a vertex operator for the tachyon field which we
write as
$$V(z,\bar z,\th,\bar\th)=T(R,\Ph_L,\Ph_R)eˆ{iq_R\Psi+iq_L\bar\Psi}
  \eqno(3.16)$$
and which is required to commute with the BRST charge given by an
expression analogous to Eq(2.10),
$$Q_{BRST}=\int dz C\left( D\P + {1\over{2k}}\J_2 \right)
 +\int d\bar z C\left(\bar D\P +{1\over{2k}}\bar{
 \J}_2\right) \eqno(3.17)$$
The zero modes of the $SL(2,R)$ currents may be represented as
differential operators acting on $T(R,\Ph_L,\Ph_R)$ of the form
$$\eqalign{\Jˆ0_2 &=  \th{\pd\over{\pd\Ph_L}} \cr\Jˆ0_\pm &=
  \th eˆ{\pm i\Ph_R}
  \left[{\pd\over{\pd R}} \mp{i\over{\sinh R}}
  \,\left({\pd\over{\pd\Ph_L}}
  -\cosh R\,{\pd\over{\pd\Ph_R}}\right)\right]}\eqno(3.18)$$
Upon requiring $V$ in Eq.(3.16) to commute with $Q_{BRST}$ in Eq.(3.17),
we find using Eqs.(3.15), (3.18) that the Virasoro operators $L_0,\bar
L_0$ are represented acting on $T(R,\Ph_L,\Ph_R)$ by the operators
$$\eqalign{L_0&=-{1\over k}\D_0(R,\Ph_L,\Ph_R) -{1\over k}
  {\pdˆ2\over{\pd\Ph_Rˆ2}} \cr \bar L_0 &= -{1\over k}\D_0(R,\Ph_L,\Ph_R)
  -{1\over k}{\pdˆ2\over{\pd\Ph_Lˆ2}}}\eqno(3.19)$$
with $\D_0$ as given by Eq.(2.20). As in Section 2, we use the physical
state condition analogous to Eq.(2.21) to decompose $T(R,\Ph_L,\Ph_R)$
as
$$\eqalign{T(R,\Ph_L,\Ph_R)&=T(R,\Ph)+\tilde T(R,\tilde\Ph) \cr
  \Ph &= {1\over 2}(\Ph_L -\Ph_R) \cr
  \tilde\Ph &= {1\over2}(\Ph_L+\Ph_R)}\eqno(3.20)$$
so that finally we find that acting on $T(R,\Ph)$, $L_0$ has the form
$$L_0=-{1\over k}\left( {\pdˆ2\over{\pd Rˆ2}} +\coth R\,
  {\pd\over{\pd R}} +{1\over4} \cothˆ2 {R\over 2}\,
  {\pdˆ2\over{\pd\Phˆ2}}\right) \eqno(3.21)$$

Identifying $L_0$ with the Laplacian derived from the tachyon effective
action, Eq.(2.18), according to Eq.(2.19), we find $g_{ij}$, $D$ given
by
$$\eqalign{dsˆ2 &= {a'k\over 4} \left( dRˆ2 +4\tanhˆ2 {R\over 2}\,
 d\Phˆ2\right)\cr D &= -\ln \cosh {R\over 2}}\eqno(3.22)$$
which we recognise as the solution Eq.(2.5) obtained by Witten in the
bosonic case by simple integration over the gauge fields. In other
words, Witten's original black hole solution is exact in the
supersymmetric case.
\vskip 1in
\line{\bigbf 4. Checking the exact solutions \hfil}
\vskip 12pt
In this section we shall check the results of the previous two sections
by showing that the proposed exact solutions in the bosonic and
supersymmetric cases are consistent with perturbative results for the
conformal invariance conditions up to four loop order. In the
supersymmetric case, it is known that there is no contribution to the
conformal invariance conditions at two [10] or three [11] loop order,
and hence four-loop calculations [12] furnish the first non-trivial
check beyond one loop. In the bosonic case, the exact solution of
Eq.(2.26) has already been checked up to three-loop order [8], but
nevertheless we feel it is worthwhile to pursue verification to the
limit of available perturbative results.

The conformal invariance conditions for the $\s$-model,
as given in Eq.(2.4), may be expressed as [18,19]
$$\eqalignno{Bˆg_{ij}&=\bˆg_{ij}+
2\nabla_{(i}S_{j)}+2\a'\nabla_i\pd_jD=0&(4.1a)
  \cr BˆD&=\bˆD+Sˆi\pd_iD+\a'\pd_iD\pdˆiD=0 &(4.1b)}$$
where $\bˆg_{ij}$, $\bˆD$ are the renormalisation group $\b$-functions
for the metric and dilaton respectively, and together with $S_i$ may be
calculated perturbatively as a power series in $\a'$. $\bˆg_{ij}$,
$\bˆD$ and $S_i$ are known up to four-loop order for both the bosonic
[20,21,22] and the $N=1$ supersymmetric [10-12] $\s$-models (and up to 5
loops for the $N=2$ supersymmetric $\s$-model [23] ). In fact, if
$\bˆg_{ij}=0$ then $\bˆD$ is guaranteed to be a constant [24], and in
our case this constant is determined by the lowest order results. Hence
we shall only need to focus our attention on Eq.(4.1a). If we postulate a
solution to Eq.(4.1) of the form
$$dsˆ2=dxˆ2+eˆ{2\l(x)}d\phˆ2,\qquad D\equiv D(x), \eqno(4.2)$$
 then we may simply
solve Eq.(4.1a) for $\l(x)$ and $D(x)$ as a power series in $\a'$. The
Christoffel symbols for the metric in Eq.(4.2) are
$$\Gˆr{}_{\ph\ph}=-\l',\qquad\Gˆ\ph{}_{r\ph}=eˆ{2\l}\l' \eqno(4.3)$$
and the Riemann tensor is given by
$$R_{klmn}={1\over2}R(g_{km}g_{ln}-g_{kn}g_{lm}) \eqno(4.4)$$
with
$$R=-2(\l''+{\l'}ˆ2).\eqno(4.5)$$
Using the bosonic results for $\bˆg_{ij}$ and $S_i$ up to four-loop order
collected in Ref. [19], we find
$$\eqalign{\l&=\l_0 +\ln t +atˆ2 +aˆ2(-3tˆ2+2tˆ4)
  \cr &\qquad +aˆ3\left\{ \left[ -{1\over6}\z(3)
  +{64\over9}\right] tˆ2 +\left[ {4\over15}\z(3)
   -{518\over45}\right] tˆ4+\left[ -{1\over10}\z(3)
  +{86\over15}\right] tˆ6\right\} +\ldots\cr
  D&= D_0 - \ln\cosh br +{1\over2}atˆ2
  +aˆ2\left[ -{1\over2}tˆ2+{1\over2}tˆ4\right]
  \cr&\qquad +aˆ3\left\{ -3          tˆ2
  +\left[ {1\over24}\z(3)+{89\over36}\right] tˆ4
  +\left[ -{1\over60}\z(3)-{13\over{30}}
  \right] tˆ6\right\} +\ldots}\eqno(4.6)$$
where $\l_0$ and $D_0$ are constants, and where
$$\eqalign{t&\equiv\tanh bx\cr a&\equiv\a' bˆ2}\eqno(4.7)$$
with $b$ a constant.
On the other hand, the proposed exact solution given in Eq.(2.26)
may be cast in the form of Eq.(4.2) by taking
$$bx\equiv{1\over2}r,\qquad a={1\over{k-2}},           \eqno(4.8)$$
and then it corresponds to taking
$$\eqalign{\l &= -\ln b +\ln t
-{1\over2}\ln\left( 1-{2a\over{1+2a}}tˆ2\right) \cr
  D&=-{1\over2}\ln  2b
  -{1\over2}\ln\sinh 2bx +{1\over2}\l (x)}\eqno(4.9)$$
which has the expansion
$$\eqalign{\l&=-\ln b +\ln t
+atˆ2 +aˆ2(-2tˆ2+tˆ4)+aˆ3(4tˆ2-2tˆ4-{2\over3}tˆ6)
+\ldots
  \cr D&=-{1\over2}\ln 4b
  -\ln\cosh bx+{1\over2}atˆ2+aˆ2(-tˆ2+{1\over2}tˆ4)
  +aˆ3(2tˆ2-tˆ4-{1\over3}tˆ6)+\ldots}\eqno(4.10)$$

Comparing Eqs.(4.6) and (4.10) we find agreement at $O({\a'}ˆ0)$
and $O(\a)$ (with appropriate choice of $\l_0$ and $D_0$ in Eq. (4.6)),
  but not at $O({\a'}ˆ2)$ or $O({\a'}ˆ3)$.
However, in the
$\s$-model approach we have the freedom to change the renormalisation
scheme used to calculate the $\b$-functions, which corresponds to making
local covariant redefinitions of the metric and dilaton fields. Hence we
should explore the possibility that Eqs. (4.6) and (4.10) are in fact
related by such a legitimate field redefinition. When considering the
effects of a field redefinition one can adopt two different (but
equivalent) viewpoints; in the first case,
which is the more usual in
the $\s$-model context, one considers the conformal invariance
conditions involving the $\b$-functions modified by a field
redefinition, and  one can then seek a solution of these new conformal
invariance conditions of the form given by Eq.(4.2). In the second case
we start from a metric and dilaton given by Eqs. (4.2),
(4.6) and simply
modify them by a local covariant redefinition. The two procedures are
manifestly equivalent, since if we redefine $g=g(\bar g)$, for
instance, we have
$$\bar\b(\bar g)=\m{d\over{d\m}}\bar g=\b(g){\pd\over{\pd g}}\bar g
  \eqno(4.11)$$
and hence $\bar\b(\bar g)=0 \iff   \b(g)=0$. The second approach is
simpler and it is the one we shall adopt. There is one subtlety,
however; a general local covariant redefinition of the metric will not
leave the metric in the form Eq.(4.2) and so we have to make a
co-ordinate change to restore the form Eq.(4.2). If we make a
redefinition
$$\eqalign{\bar g_{ij}&=g_{ij}+T_{ij}\cr\bar D&=D+K}\eqno(4.12)$$
(where $T_{ij}$ and $K$ are restricted to be local covariant quantities)
then to recover the form Eq.(4.2) we also need to make a co-ordinate
transformation
$$\tilde x=x+q(x)\eqno(4.13)$$
where
$$q' ={1\over2}T_{xx}\eqno(4.14)$$
so that the resulting combined field redefinition and co-ordinate
transformation of $\l$, $D$ is
$$\eqalign{\tilde\l &= \l-q\l'+{1\over2}eˆ{-2\l}T_{\ph\ph}\cr
  \tilde D &=  D-qD'   +K}\eqno(4.15)$$

{}From the other point of view, in the case of a
$\s$-model with a target space of arbitrary dimensions, at three and
higher loops there are tensor structures in the $\b$-functions which are
\ss\ss scheme-independent" and cannot be modified by field redefinitions
[25].  However in the present case of a two dimensional target space,
many curvature invariants which are in general distinct become related,
and one might expect that there would no longer be any invariant
structures. Nevertheless, a remnant of scheme independence persists,
since it is not possible to make an arbitrary redefinition of $\l$ via
Eq.(4.15). In fact, an $O({\a'}ˆn)$ covariant metric redefinition of the
form Eq.(4.12) makes a change in $\l$ of the form
$$\d\l=\sum_{i=1}ˆn\l_itˆ{2i}\eqno(4.16)$$
with the constraint
$$\sum \l_i=0\eqno(4.17)$$

Comparing Eqs.(4.6) and (4.10), we see that the difference between the
respective solutions for $\l$ and $D$ is given by
$$\eqalign{\d\l&=aˆ2(-tˆ2+tˆ4)+aˆ3 \cr
 &\qquad +\left\{\left( -{1\over6}\z(3)+{28\over9}\right) tˆ2
  +\left( {4\over15}\z(3)-{428\over45}\right) tˆ4
         +\left( -{1\over10}\z(3)+{32\over5}\right) tˆ6
  \right\}+\ldots \cr \d D&=aˆ2\left( {1\over2}tˆ2 \right)
  +aˆ3\left\{ -5          tˆ2 +\left( {1\over24}\z(3)
  +{125\over36}\right) tˆ4
  +\left( -{1\over60}\z(3)-{1\over10}\right) tˆ6\right\}+\ldots \cr}
  \eqno(4.18)$$
and it is clear that the constraint Eq.(4.17) is satisfied at
$O({\a'}ˆ2)$ and $O({\a'}ˆ3)$.
This  analysis was first carried out at
$O({\a'}ˆ2)$ in Ref. [8] where an explicit local covariant field
redefinition was displayed which yields the correct $\d\l$ and $\d D$ at
this order. At $O(\a'ˆ3)$, the most general local field redefinition is
obtained by taking in Eq. (4.12)
$$\eqalign  {T_{ij}&=a_1Rˆ3g_{ij}+a_2\pd_iR\pd_jR+a_3\nabla_i\pd_jR
+a_4\pd R.\pd R\cr
&\qquad+a_5R\nablaˆ2Rg_{ij}+a_6\nablaˆ2\nablaˆ2Rg_{ij}
+a_7\nabla_i\nabla_j\nablaˆ2R\cr
K&=b_1Rˆ3+b_2\pd R.\pd R+b_3R\nablaˆ2R+b_4\nablaˆ2\nablaˆ2R.       \cr}
\eqno(4.19)$$
There is a certain amount of freedom in selecting a field redefinition
which effects the required transformation; one possibility is
$$\eqalign  {a_1={1\over{256}}\z(3)+{5\over{48}},&\qquad a_2=a_3=0,\cr
a_4={17\over{48}},\qquad a_5=a_6&=a_7=0,\qquad b_1={1\over{1536}}
(\z(3)-205),\cr
b_2=-{77\over{768}},&\qquad b_3=b_4=0.\cr}\eqno(4.20)$$
An interesting phenomenon is that the coefficients of the terms in
Eq. (4.19) which involve total derivatives, namely $a_6$, $a_7$ and
$b_4$, turn out to be related amongst themselves by a homogeneous
equation
$$4b_4=2a_6-a_7,\eqno(4.21)$$
which means that $a_6$, $a_7$ and $b_4$ can be chosen to be all zero,
independently of the values assigned to the other coefficients. The
same effect can also be observed at the three-loop order, although its
significance is still unclear.

Hence we have shown that up to $O(\a'ˆ3       )$, corresponding to
four-loop order, the exact black hole
solution of Eq.(2.26) is a solution of the $\s$-model conformal
invariance conditions in some renormalisation scheme. In fact, the
validity at this order of Witten's original solution Eq.(2.5) in the
$N=1$
supersymmetric case  follows as an immediate corollary. The
$\b$-function for the $N=1$ supersymmetric $\s$-model is the same as in
the bosonic case
at one loop, zero at two and three loops, and at four loops
can be obtained from the four-loop bosonic $\s$-model $\b$-function by
retaining only the $\z(3)$ terms [21]. The $\z(3)$ terms in Eq.(4.6),
and hence Eq.(4.18), are therefore unchanged in passing from the bosonic
to the supersymmetric case, and the analogue of Eq.(4.6) in the
supersymmetric case consists of the zeroth order term and the $\z(3)$
terms. Since the zeroth order term in Eq.(4.6) reproduces Witten's
original solution Eq.(2.5), and since the $\z(3)$ terms in Eq.(4.6)
have the property Eq.(4.17), it follows that the Witten solution
Eq.(2.5) is valid up to four loops in the $N=1$ supersymmetric case. In
fact, since the $N=1$ supersymmetric WZW model is known to possess $N=2$
supersymmetry [26], and since the five-loop contribution to the
$\b$-function for the $N=2$ supersymmetric $\b$-function can be
field-redefined to zero [23], we can deduce that the Witten solution
is valid up to five-loop order.
\vskip 1in
\line{\bigbf Conclusions \hfil}
\vskip 12pt
We have proved that Witten's original black-hole metric, Eq.(2.5), is
exactly conformally invariant in the supersymmetric case. We have
demonstrated this by explicit comparison with the $\s$-model conformal
invariance conditions up to five-loop order. We have also shown that
the exact solution for the bosonic case, proposed in Ref. [7] and given
in Eq. (2.26), is consistent with the $\s$-model conformal invariance
conditions up to four-loop order.

Finally, a comment is in order on the derivation of the exact solution
in [7]. This derivation relies on identifying the Virasoro operators
$L_0$, $\bar L_0$ with the one-loop string effective action for the
tachyon field. However, the tachyon string effective action gets
contributions at three and higher orders in perturbation theory in
minimal subtraction. Nevertheless, if one included these contributions
the effect would simply be a further local covariant redefinition of the
metric and hence could not be detected by the methods used here.
To this extent, there is an intrinsic ambiguity in the identification of
the metric and dilaton fields starting from the gauged WZW model.
\vskip 1in
\line {\bigbf Acknowledgements \hfil}
\vskip 12pt
We are indebted to Hugh Osborn and Arkady Tseytlin for useful and
illuminating discussions. I. J. and J. P. thank the S. E. R. C. for
support.
\vskip 1in
\line {\bigbf References\hfil}
\vskip 12pt
\item {1. } E. Witten, Phys. Rev. {\bf D44} (1991) 314.
\item {2. } C. G. Callan, R. C. Myers and M. J. Perry, Nucl. Phys.
{\bf B311} (1988) 673;
\item {   } G. W. Gibbons and K. Maeda, Nucl. Phys. {\bf B298} (1988) 741
 ;
\item {   } D. Garfinkle, G. T. Horowitz and A. Strominger, Phys. Rev.
{\bf D43} (1991) 3140;
\item {   } G. T. Horowitz and A. Strominger, Nucl. Phys. {\bf B360}
(1991) 197.
\item {   } G. Mandal, A. Sengupta and S. Wadia, Mod. Phys. Lett.
{\bf A6} (1991) 1685.
\item {3. } N. Ishibashi, M. Li and A. R. Steif,
``Two-dimensional charged
black holes in string theory'', Santa Barbara preprint UCSBTH-91-28;
\item {   } J. H. Horne and G. T. Horowitz,
 ``Exact black string solutions
in three dimensions'', Santa Barbara preprint UCSBTH-91-39;
\item {   } P. Horava, ``Some exact solutions of string theory in four
and five dimensions'', Enrico Fermi Institute preprint EFI-91-57;
\item {   } S. P. Khastgir and A. Kumar, ``String effective action and
two-dimensional charged black hole'',
            Bhubaneswar preprint IP/BBSR/91-31;
\item {   } M. D. Mc.Guigan, C. R. Nappi and S. A. Yost, ``Charged
black holes in two-dimensional string theory'', preprint IASSNS-HEP-91/57;
\item {   } M. Gasperini, J. Maharana and G. Veneziano,
``From trivial to
non-trivial conformal string backgrounds via $O(d,d)$ transformations'',
preprint CERN-TH-6214/91.
\item {4. } E. Witten, Comm. Math. Phys. {\bf 92} (1984) 455.
\item {5. } P. di Vecchia and P. Rossi, Phys. Lett. {\bf B140} (1984)
334;
\item {   } P. di Vecchia, B. Durhuus and J. L. Petersen, Phys. Lett.
{\bf B144} (1984) 245;
\item {   } D. Gonzales and A. N. Redlich, Phys. Lett. {\bf B147} (1984)
150;
A. N. Redlich and H. J. Schnitzer, Phys. Lett. {\bf B193} (1987) 471.
\item {6. } P. Goddard, A. Kent and D. Olive, Phys. Lett. {\bf B152}
(1985) 88;
\item {   } K. Gaw\c edski and A. Kupiainen, Phys. Lett. {\bf B215} (1988)
119;
\item {   } Nucl. Phys {\bf B320} (1989) 625.
\item {7. } R. Dijkgraaf, E. Verlinde and H. Verlinde, ``String
propagation in a black hole geometry'', preprint IASSNS-HEP-91/22.
\item {8. } A. A. Tseytlin, Phys. Lett. {\bf B268} (1991) 175.
\item {9. } P. di Vecchia, V. G. Knizhnik, J. L. Petersen and P. Rossi,
Nucl. Phys. {\bf B253} (1985) 701.
\item {10.} L. Alvarez-Gaum\'e, D. Z. Freedman and S. Mukhi, Ann. Phys.
(N. Y.) {\bf 134} (1981) 85.
\item {11.} L. Alvarez-Gaum\' e, Nucl. Phys. {\bf B184} (1981) 180.
\item {12.} M. T. Grisaru, A. E. M. van de Ven and D. Zanon, Nucl. Phys.
{\bf B277} (1986) 409.
\item {13.} C. M. Hull and B. Spence, Phys. Lett. {\bf B232} (1989) 204;
Nucl. Phys. {\bf B353} (1991) 379; Mod. Phys. Lett. {\bf A6} (1991) 969;
Nucl. Phys. {\bf B363} (1991) 593;
\item {   } I. Jack, D. R. T. Jones, N. Mohammedi and H. Osborn,
Nucl. Phys. {\bf B332} (1990) 359.
\item {14.} A. Giveon, Mod. Phys. Lett. {\bf A6} (1991) 2843;
\item {   } E. B. Kiritsis, Mod. Phys. Lett. {\bf A6} (1991) 2871;
\item {   } M. Ro\v cek and E. Verlinde, ``Duality, quotients and
currents'', preprint IASSNS-HEP-91/68.
\item {15.} See P. Goddard and D. Olive, Int. J. Mod. Phys. {\bf A1}
(1986) 303
            for a review and further references.
\item {16.} S. Nojiri, ``Superstring in two-dimensional black hole'',
preprint FERMILAB-PUB-91/230-T.
\item {17.} J. Fuchs, Nucl. Phys. {\bf B286} (1986) 455;
\item {   } C. M. Hull and B. Spence, Phys. Lett. {\bf B241} (1990) 357.
\item {18.} G. M. Shore, Nucl. Phys. {\bf B286} (1987) 349.
\item {19.} A. A. Tseytlin, Phys. Lett. {\bf B178} (1986) 34; Nucl Phys.
{\bf B294} (1987) 383.
\item {20.} D. Friedan, Phys. Rev. Lett. {\bf 51} (1980) 334; Ann. Phys.
(N. Y.) {\bf 163} (1985) 316;
\item {   } A. A. Tseytlin, Nucl. Phys. {\bf B276} (1986) 391;
\item {   } S. J. Graham, Phys. Lett. {\bf B197} (1987) 543;
\item {   } A. P. Foakes and N. Mohammedi, Phys. Lett. {\bf B198} (1987)
359; Nucl. Phys. {\bf B306} (1988) 343;
\item {   } I. Jack, D. R. T. Jones and D. A. Ross, Nucl. Phys.
{\bf B307} (1988) 531.
\item {21.} I. Jack, D. R. T. Jones and N. Mohammedi, Nucl. Phys.
{\bf B322} (1989) 431.
\item {22.} I. Jack, D. R. T. Jones and N. Mohammedi, Nucl. Phys.
{\bf B332} (1990) 330.
\item {23.} M. T. Grisaru, D. I. Kazakov and D Zanon, Nucl. Phys.
{\bf B287} (1987) 189.
\item {24.} G. Curci and G. Paffuti, Nucl. Phys. {\bf B286} (1987) 399.
\item {25.} R. R. Metsaev and A. A. Tseytlin, Phys. Lett. {\bf B191}
(1987) 354.
\item {26.} Y. Kazama and H. Suzuki, Nucl. Phys. {\bf B321} (1989) 232.
\vfill
\eject
\end